\documentclass[aps,prd,twocolumn,showpacs,preprintnumbers,nofootinbib]{revtex4}
\usepackage[dvips]{graphicx}
\usepackage{bm,latexsym,amsmath,amssymb,amsfonts}
\newcommand*{\car}{{\cal R}}
\newcommand*{\cah}{{\cal H}}
\newcommand*{\cag}{{\cal G}}
\begin{document}

\title{
Five-dimensional black strings in Einstein-Gauss-Bonnet gravity
}

\author{Tsutomu~Kobayashi}
\email{tsutomu@tap.scphys.kyoto-u.ac.jp}
\author{Takahiro~Tanaka}
\email{tama@scphys.kyoto-u.ac.jp}

\affiliation{
Department of Physics, Kyoto University, Kyoto 606-8502, Japan 
}


\begin{abstract}
We consider black-string-type solutions
in five-dimensional Einstein-Gauss-Bonnet gravity.
Numerically constructed solutions under static, axially symmetric 
and translationally invariant metric ansatz are presented. 
The solutions are specified by two asymptotic charges:
mass of a black string and a scalar charge
associated with the radion part of the metric. 
Regular black string solutions are found if and only if the two charges
satisfy a fine-tuned relation,
and otherwise the spacetime develops 
a singular event horizon or a naked singularity.
We can also generate bubble solutions from 
the black strings by using a double Wick rotation.
\end{abstract}


\preprint{KUNS-1954}

\maketitle

\section{Introduction}

Recently there has been a growing attention to
gravity in higher dimensions,
and one of the central interests
is directed toward
understanding the nature of higher dimensional black objects.
The terminology ``black objects'' stems from the curious fact that
in $d (\geq 5)$ dimensions
various horizon topologies are allowed~\cite{Emparan:2001wn}
as well as a spherical topology $S^{d-2}$~\cite{Myers:1986un}.
For example, there are stringlike solutions
in five dimensions,
which are constructed by simply extending
four-dimensional black holes into one extra dimension,
and thus they have horizon topology $S^2 \times  {\rm R}$.
Black string solutions which are not uniform in the direction 
of an extra dimension were also found 
numerically~\cite{Wiseman:2002zc,Kudoh:2004hs}. 
Another example is a five-dimensional rotating black ring
with horizon topology $S^2\times S^1$~\cite{Emparan:2001wn}.
This solution gives clear evidence against 
black hole uniqueness in higher dimensions.
As for stability, 
while it is well known that a uniform black string suffers
from the Gregory-Laflamme instability at long wavelengths~\cite{Gregory:1993vy},
stability of the other black solutions 
has not been fully understood yet and
is now being studied actively~\cite{Kodama:2003jz,Ishibashi:2003ap,Kodama:2003kk,Emparan:2003sy,Dotti:2004sh}.

Motivated by the suggestion from particle physics that our observable world
may be a brane embedded in a higher dimensional bulk spacetime,
configurations composed of a black object-brane system
have also been an interesting issue in the past few years.
A black string solution in the Randall-Sundrum braneworld
was first studied in~\cite{Chamblin:1999by}.
This solution is an extension of
the Schwarzschild black string in flat space
to a solution in an Anti-de Sitter
bulk bounded by a brane with a tuned tension. 
Much more nontrivial is a black hole localized on
a brane.
No exact solutions describing such black holes
have been known so far,
but numerical solutions were worked out in~\cite{Kudoh:2003xz},
and an analytic approximation method was developed in~\cite{Karasik:2003tx}.
Black holes in braneworld scenarios have
a phenomenological aspect as well.
For instance, an exciting possibility was pointed out that
black holes may be produced at accelerators~\cite{Giddings:2001bu, Dimopoulos:2001hw},
and a characteristic gravitational wave signal from a black string
was discussed~\cite{Seahra:2004fg}.

In this paper we investigate black string solutions
of five-dimensional Einstein gravity with
a ``Gauss-Bonnet'' higher curvature correction term.
The Gauss-Bonnet combination of curvature invariants
is relevant in five dimensions or higher, 
although it reduces to just a total derivative in four dimensions.
The Gauss-Bonnet term is of importance in string theory
since it arises as the next-to-leading order
correction in the heterotic string effective action.
This fact motivates several past
studies~\cite{Mignemi:1992nt,Kanti:1995vq,Alexeev:1996vs,Torii:1996yi}
which considered black hole solutions
in a gravitational theory with a Gauss-Bonnet correction term.
(Though they have focused on a \textit{four}-dimensional case,
it does not contradict the above remark since in these works
a scalar field coupled dilatonically to the Gauss-Bonnet term 
is introduced.)
Very recently the Gauss-Bonnet correction
is frequently discussed in the braneworld context
as an attempt to extend the Randall-Sundrum model
in a natural way~\cite{gaussbonnet, Charmousis:2003ke}.
Five-dimensional,
static and spherically symmetric black holes
were systematically examined
in the presence of a Gauss-Bonnet term~\cite{gbbh, Konoplya:2004xx, Cvetic:2001bk, Charmousis:2002rc},
and its consequences on the brane cosmology were
discussed~\cite{Charmousis:2002rc, Nojiri:2000gv, gbcosmology}.
Our study here addresses the simplest setup:
the Einstein-Hilbert plus the Gauss-Bonnet
curvature term in five dimensions
without a cosmological constant or branes.


The structure of the present paper is as follows.
In the next section we outline Einstein-Gauss-Bonnet gravity
and we consider static, axially symmetric, and translationally invariant
black string metric ansatz and its asymptotic boundary conditions
in the framework of Einstein-Gauss-Bonnet gravity.
In Sec.~III we briefly discuss our problem
from a viewpoint of the reduced theory in four dimensions.
In Sec.~IV, first we provide an approximate analytic solution
valid for large mass, and then present our numerical results.
We also discuss bubble solutions generated from the black string solutions.
Section V contains some remarks.

\section{Preliminaries}

\subsection{Einstein-Gauss-Bonnet gravity}

The gravitational theory that we consider
in this paper is defined by the action
\begin{eqnarray}
{\cal S}={1\over 16\pi G_5} \int d^5x \sqrt{-g}[\car
+\alpha {\cal L}_{GB}],
\label{5dGBaction}
\end{eqnarray}
where $G_5$ is the five-dimensional Newton's constant,
and the Gauss-Bonnet Lagrangian is given by
\begin{eqnarray}
{\cal L}_{GB}= \car^2-4\car_{cd}\car^{cd}+\car_{cdef}\car^{cdef}. 
\label{GB_lagrangian}
\end{eqnarray}
A parameter $\alpha$ has dimension of (length)$^2$.  
Variation of this action gives following field equations
for the metric:
\begin{eqnarray}
\cag_{ab}=\frac{\alpha}{2}{\cal H}_{ab},
\label{field_equations}
\end{eqnarray}
where $\cag_{ab}$ is the Einstein tensor and
$\cah_{ab}$ is the Lanczos tensor,
\begin{eqnarray}
&&\cah_{ab}=
{\cal L}_{GB}g_{ab}
-4(
\car\car_{ab}-2\car_{ac}\car^c_{~b}
\nonumber\\&&\qquad\qquad\qquad
-2\car_{acbd}\car^{cd}+\car_{acde}\car_b^{~cde}
).
\end{eqnarray}
Although the action involves the terms higher order in curvature,
the field equations~(\ref{field_equations}) are second order.
This is the special feature of the Gauss-Bonnet Lagrangian~(\ref{GB_lagrangian}). 
For a nice review on Einstein-Gauss-Bonnet gravity
(and its Lovelock generalization),
see, e.g.,~\cite{Deruelle:2003ck}.

\subsection{Two coordinate systems for black string spacetime}

We are interested in a black-string-type solution of
the field equations~(\ref{field_equations}).
We write a general metric of
a static black string spacetime as
\begin{eqnarray}
ds^2=
-e^{2A(r)}dt^2+e^{2B(r)}dr^2+r^2d\Omega^2+e^{2\Phi(r)}dy^2,
\label{r_coordinate}
\end{eqnarray}
where $d\Omega^2=d\theta^2 + \sin^2\theta d\varphi^2$
is the line element on a unit 2-sphere,
$r$ is the circumferential radius
and $y$ is a coordinate in the direction of an extra dimension.
A straightforward calculation gives 
explicit forms of the Einstein and Lanczos tensor components,
which are presented in Appendix A.

As pointed out in \cite{Barcelo:2002wz},
the radion part $g_{yy}=e^{2\Phi(r)}$ is crucial for
constructing a black string solution
in the framework of Einstein-Gauss-Bonnet gravity.
In fact, if we assume $\Phi(r)=$ const,
then it follows from the $(t,t)$ and $(r,r)$
components of the field equations
that $e^{-2B(r)}=e^{2A(r)}=1-2m/r$
where $m$ is an integration constant.
This is incompatible with the $(y,y)$ component
of the field equations unless $\alpha =0$.

The event horizon of a black string is
located at some finite radius in the coordinate system
of~(\ref{r_coordinate}), $e^{2A(r_h)}=0$.
Such singularity in the metric component might cause
difficulty when integrating the field equations numerically,
and thus
it will be better to use the so-called ``tortoise'' coordinate
defined by $d\rho^2 = e^{2B(r)-2A(r)}dr^2$.
Using this coordinate system,
the metric is expressed as
\begin{eqnarray}
ds^2=N^2(\rho)(-dt^2+d\rho^2)+r^2 (\rho)d\Omega^2+C^2(\rho)dy^2.
\label{kame}
\end{eqnarray}
Here we write $e^{A(r)}=N(\rho)$ and $e^{\Phi(r)}=C(\rho)$.
The Einstein and Lanczos tensors in this coordinate system
are also presented in Appendix A.
Now the horizon $r=r_h$ is mapped to $\rho=-\infty$,
and the metric components are expected to be 
slowly varying functions of $\rho$ in this limit, 
which makes a numerical investigation easier.
Also, there is another advantage of using this coordinate system. 
As will be seen,
$r(\rho)$ is not a monotonic function for a class of solutions.
Due to this unusual behavior, $g_{tt}$ and $g_{yy}$
are two-valued functions of $r$, while they are still single-valued in $\rho$,
and hence $\rho$ is an appropriate coordinate for
describing such a class of solutions.

\subsection{Asymptotic boundary conditions}

Let us investigate the asymptotic behavior of the metric functions.
First we work in the coordinate system of~(\ref{r_coordinate}).
For sufficiently large $r$, we may neglect ${\cal H}_a^{~b}$
because it asymptotically decays in higher powers of $1/r$, and
the field equations become ${\cal G}_a^{~b}\approx 0$.
Then we have
\begin{eqnarray}
&&\quad {\cal G}_t^{~t}+{\cal G}_{r}^{~r}+{\cal G}_{\theta}^{~\theta}
+{\cal G}_{\varphi}^{~\varphi}-2{\cal G}_y^{~y}=
\nonumber\\
&&3e^{-2B}\left[
\Phi''+(\Phi')^2+\left(A'-B'+\frac{2}{r}\right)\Phi' \right]\approx 0,
\quad
\label{eq_for_Phi}
\end{eqnarray}
where the prime denotes differentiation with respect to $r$.
This equation can be integrated to give
\begin{eqnarray}
&&{\left(e^{\Phi}\right)}' \approx -Q\frac{e^{B-A}}{r^2},
\label{scalar_charge}\\
&&e^{\Phi}\approx 1+Q\int^{\infty}_{r}\frac{e^{B-A}}{r^2}dr
\approx1+\frac{Q}{r},
\label{charge_and_phi}
\end{eqnarray}
where $Q$ is an integration constant.
From Eq.~(\ref{charge_and_phi}), we obtain
\begin{eqnarray}
-\frac{1}{4\pi} \int \partial_r \Phi \cdot r^2 d\Omega=Q,
\end{eqnarray}
where the integral is over a 2-sphere at infinity.
Thus, $Q$ is regarded as a charge associated with
the ``scalar field'' $\Phi$, which we called radion earlier.

When the Gauss-Bonnet term is turned off ($\alpha=0$),
the relation~(\ref{scalar_charge}) is \textit{exact}
for all values of $r$.
Then, it can be easily seen that
$(e^{\Phi})'$ and $e^{\Phi}$ diverge at the horizon, $r=r_h$,
because $e^{A(r_h)}=0$.
To avoid this, we must set $Q=0$, and hence
a black string solution in Einstein gravity cannot possess
a ``scalar charge''.
In Einstein-Gauss-Bonnet gravity, however,
there is no reason for prohibiting a nonzero scalar charge.

Another set of field equations yields
\begin{eqnarray}
&&\quad 2{\cal G}_{\theta}^{~\theta}+{\cal G}_{r}^{~r}-{\cal G}_t^{~t}
\nonumber\\
&&=2e^{-2B}\left[
A''+(A')^2+\left(\Phi'-B'+\frac{2}{r}\right)A' 
\right]\approx 0,\quad
\end{eqnarray}
where we have used Eq.~(\ref{eq_for_Phi}).
From this we similarly obtain
\begin{eqnarray}
&&{\left(e^A\right)}' \approx M\frac{e^{B-\Phi}}{r^2},
\\
&&e^{A}\approx 1-M\int^{\infty}_{r}\frac{e^{B-\Phi}}{r^2}dr
\approx 1-\frac{M}{r},
\label{asy_M}
\end{eqnarray}
where $M$ is an integration constant.
Since $-g_{tt}\approx 1-2M/r$ near infinity,
$M$ is indeed the Arnowitt-Deser-Misner mass.
Finally, the asymptotic behavior of $g_{rr}$ is determined
by the equation $\cag_r^{~r}\approx 0$
as
\begin{eqnarray}
e^{2B(r)}\approx 1+\frac{2(M-Q)}{r}.
\end{eqnarray}
Thus the asymptotic form of the metric functions is
specified by two parameters, $M$ and $Q$.

Now let us move on to the second coordinate system~(\ref{kame}),
which we will use for actual numerical calculations.
Asymptotic forms of $N(\rho)$ and $C(\rho)$
are, of course,
\begin{eqnarray}
&&N(\rho) \approx 1-\frac{M}{r(\rho)},
\label{boundary1}
\\
&&C(\rho) \approx 1+\frac{Q}{r(\rho)}.
\label{boundary2}
\end{eqnarray}
Since $r'=e^{A-B}$ by definition, 
we have for a large value of $\rho$ 
\begin{eqnarray}
r'(\rho)\approx 1-\frac{2M}{r(\rho)} +\frac{Q}{r(\rho)}.
\label{boundary3}
\end{eqnarray}

Although Einstein-Gauss-Bonnet gravity
has a parameter of dimension of (length)$^2$, $\alpha$,
it does not appear in the above asymptotic analysis.
This parameter can always be set equal to unity
in the field equations (provided that $\alpha > 0$)
by redefining the radial coordinate as $r\to r/\alpha^{1/2}$
(or $\rho \to \rho/\alpha^{1/2}$)
and the asymptotic charges as $M\to M/\alpha^{1/2}$,
$Q\to Q/\alpha^{1/2}$.
In this sense, free parameters that specify
black string solutions are two dimensionless combinations,
$M/\alpha^{1/2}$ and $Q/\alpha^{1/2}$.

\section{Four-dimensional point of view}

Translational invariance in the $y$ direction
of black string spacetime 
allows us to analyze such spacetime from
a point of view of the reduced theory in four dimensions.
Before proceeding to solving the field equations,
let us take a brief look at the problem from this viewpoint.

Substituting the metric ansatz
\begin{eqnarray}
g_{ab}dx^adx^b=e^{-\Phi(\hat x)}
q_{\mu\nu}(\hat x)d\hat x^{\mu}d\hat x^{\nu}
+e^{2\Phi(\hat x)}dy^2,
\label{einframe}
\end{eqnarray}
into the five-dimensional action~(\ref{5dGBaction}),
we obtain the following four-dimensional action:
\begin{eqnarray}
&&S = \frac{1}{16\pi G_4}\int d\hat x^4\sqrt{-q}
\bigg\{ R - \frac{3}{2}(\nabla \Phi)^2 +\alpha
e^{\Phi} L_{GB}
\nonumber\\
&&+\alpha e^{\Phi}[
4G_{\mu\nu}\nabla^{\mu}\Phi\nabla^{\nu}\Phi
-3\Box\Phi(\nabla \Phi)^2
]\bigg\},
\label{reduced_action}
\end{eqnarray}
where $G_4$ is the four-dimensional Newton's constant,
$R$, $G_{\mu\nu}$, and $L_{GB}$
are constructed from the metric $q_{\mu\nu}$,
$\nabla_{\mu}$ is the derivative operator
associated with $q_{\mu\nu}$, and
we have used a notation 
$(\nabla \Phi)^2:=\nabla_{\mu}\Phi \nabla^{\mu}\Phi$,
$\Box\Phi:= \nabla_{\mu}\nabla^{\mu}\Phi$.
With some algebra
we have omitted the total derivative terms in Eq.~(\ref{reduced_action}).

The reduced action consists of the Einstein-Hilbert term and
a scalar field $\Phi$,
which has a non-trivial kinetic term and is coupled
to the four-dimensional Gauss-Bonnet term
as well as to the Einstein tensor.
Unfortunately, this action itself
does not seem to bring us much insight 
on the properties of the black strings.
However, black hole solutions of the rather simpler action
\begin{eqnarray}
S = \frac{1}{16\pi G_4}\int d x^4\sqrt{- q}
\left[ R - \frac{3}{2}(\nabla \Phi)^2 +\alpha e^{\Phi}L_{GB}
\right],
\label{dilaton_GB}
\end{eqnarray}
which contains only the terms in the first line in the action~(\ref{reduced_action}),
are extensively studied in
Refs.~\cite{Mignemi:1992nt,Kanti:1995vq,Alexeev:1996vs,Torii:1996yi}.
It is worth noting beforehand that
our black string solutions share some of the properties
with four-dimensional black hole solutions of the
action~(\ref{dilaton_GB}).
For example, the same exotic structure of the metric near $r=r_m$
(which is to be mentioned in the next section)
is addressed in Ref.~\cite{Kanti:1995vq}.

\section{Black string solutions in Einstein-Gauss-Bonnet gravity}

\subsection{Perturbative solution}

Before presenting numerical results,
we shall construct a regular black string solution analytically
by taking the Gauss-Bonnet term as a perturbation
from pure Einstein gravity.

We start from the Schwarzschild black string in Einstein gravity,
\begin{eqnarray}
ds^2_0=-f(r) dt^2
+f^{-1}(r)dr^2+r^2d\Omega^2+dy^2,
\label{Sch_st}
\end{eqnarray}
with $f(r):=1-2M/r$.
We solve the field equations iteratively by
taking $\alpha$ as a small expansion parameter around this background.
Substituting the background solution~(\ref{Sch_st}) into the Lanczos tensor,
we have
\begin{eqnarray}
{\cal H}_{y}^{~y}=\frac{48M^2}{r^6},\quad
{\cal H}_{t}^{~t}={\cal H}_{r}^{~r}={\cal H}_{\theta}^{~\theta}={\cal H}_{\varphi}^{~\varphi}=0.
\end{eqnarray}
We make the ansatz
\begin{eqnarray}
&&e^{2A(r)}=f(r) +2\alpha A_1(r)+ \alpha^2 A_2(r)+\cdots,
\\
&&e^{-2B(r)}=f(r)-2\alpha B_1(r)- \alpha^2 B_2(r)+\cdots,
\\
&&e^{\Phi(r)}=1+\alpha \Phi_1(r) +\alpha^2 \Phi_2(r) +\cdots,
\end{eqnarray}
and then the equation for $\Phi_1(r)$ is obtained as
\begin{eqnarray}
\left[r(r-2M) {\Phi_1}' \right]'=-16\frac{M^2}{r^4}.
\end{eqnarray}
A solution which vanishes at infinity and is regular 
at the horizon is given by
\begin{eqnarray}
\Phi_1(r) = \frac{2}{3}\frac{1}{M}\left(
\frac{1}{r}+\frac{M}{r^2}+\frac{4M^2}{3r^3} \right).
\label{1st_order_phi}
\end{eqnarray}
Comparing this with Eq.~(\ref{charge_and_phi}),
we see that the asymptotic charge
$Q$ is not independent of mass $M$ for this solution: 
\begin{eqnarray}
Q=\frac{2\alpha}{3M}+{\cal O}(\alpha^2).
\label{relation_Q_and_M}
\end{eqnarray}
The other two sets of the field equations of ${\cal O}(\alpha)$ give
\begin{eqnarray}
&&A_1(r)=\frac{2}{3}\left(\frac{1}{r^2}+\frac{5M}{6r^3}+\frac{2M^2}{r^4} \right),
\\
&&B_1(r)=- \frac{2}{3M}\frac{1}{r}-\frac{1}{3}
\left(\frac{1}{r^2}+\frac{M}{r^3}-\frac{20M^2}{3r^4} \right).
\end{eqnarray}
Here one of the two integration constants is
determined by the condition that $A_1$ and $B_1$ vanish at infinity,
and the other, which appears in the form of
$+c/r$ (respectively, $-c/r$) in $A_1$ (respectively, $B_1$),
is set to be zero since we are considering a solution
with a fixed mass $M$.
The event horizon is located at
\begin{eqnarray}
r_h=2M-\frac{23}{18}\frac{\alpha}{M} + {\cal O}(\alpha^2),
\end{eqnarray}
with which it can be seen that $e^{2A(r_h)}=e^{-2B(r_h)}={\cal O}(\alpha^2)$.

A comment is now in order.
Mignemi and Stewart~\cite{Mignemi:1992nt}
investigated black hole solutions
of the action~(\ref{dilaton_GB})
in the same way as above.
Namely, they
perturbatively obtained a solution valid at first order in $\alpha$,
assuming the Schwarzschild black hole solution to be a background.
Their solution for the scalar field coincides with
ours~(\ref{1st_order_phi}) up to the overall factor
[see Eq.~(5) of Ref.~\cite{Mignemi:1992nt}].
This is not a surprise because
$G_{\mu\nu}$ and $\nabla_{\mu}\Phi$ vanish for the background solution
and hence the second line of the action~(\ref{reduced_action}) 
does not contribute to the quadratic order of perturbation.

\subsection{Numerical results}

The numerical integration is performed by
using the coordinate system~(\ref{kame}).
For given values of $M$ and $Q$,
we impose the asymptotic boundary conditions
following Eqs.~(\ref{boundary1})--(\ref{boundary3}),
and then integrate the $(t,t)$, $(\theta,\theta)$, and $(y,y)$ components of the
field equations starting from sufficiently large $\rho$ towards $\rho\to-\infty$.
The constraint equation
[the $(r,r)$ component] is used to make sure that numerical errors are 
sufficiently small.

\begin{figure}[t]
  \begin{center}
    \includegraphics[keepaspectratio=true,height=160mm]{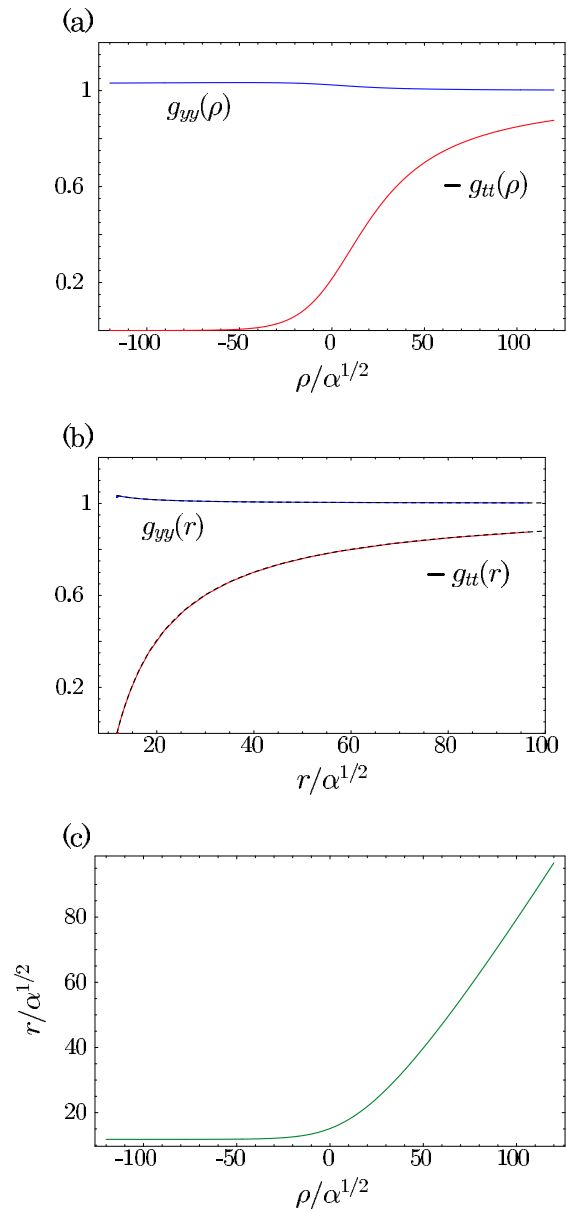}
  \end{center}
  \caption{(color online)
  Metric components $g_{tt}$ (red lines) and $g_{yy}$ (blue lines)
  as functions of the tortoise coordinate $\rho$ (a)
  and the circumferential radius $r$ (b),
  and the relation between $\rho$ and $r$ (c)
  for a black string with rather large mass
  $M=6\alpha^{1/2}$ and $Q=2\alpha/3M\approx Q_c$.
  The two parameters are fine-tuned so that
  the solution shows a regular behavior.
  The perturbative solution is plotted as a reference [dashed line in (b)].
  }
  \label{fig:regular.eps}
\end{figure}

\paragraph{Regular black string}
If and only if a relation between $M$ and $Q$ is fine-tuned, we can obtain
a regular black string solution, an example of which is 
presented in Fig.~\ref{fig:regular.eps}.
For a large mass black string, the fine-tuned relation
$Q=Q_c(M)$ is
well approximated by Eq.~(\ref{relation_Q_and_M}),
\begin{eqnarray}
Q_c(M) \simeq \frac{2\alpha}{3M}, \quad M^2\gg\alpha,
\end{eqnarray}
and the solution is of course well approximated by the perturbative solution
given in the previous subsection.
On the other hand, for smaller $M$
the regular solutions deviate from the perturbative solution
due to the higher-order terms in $\alpha$,
though their effect is not so large.
The circumferential radius $r$ is a monotonic function of $\rho$
for this class of solutions.

This solution has a regular event horizon,
and near the horizon the metric components
are expressed as~\cite{Kanti:1995vq, Alexeev:1996vs}
\begin{eqnarray}
&&e^{2A(r)}=a_1x+\frac{a_2}{2}x^2+\cdots,\\
&&e^{-2B(r)}=b_1x+\frac{b_2}{2}x^2+\cdots,\\
&&\Phi(r)=\Phi_h +\phi_1x+\frac{\phi_2}{2}x^2+\cdots,
\end{eqnarray}
where $x=r-r_h$.
Substituting these into the field equations
and solving the resultant algebraic equations,
we can determine the coefficients order by order.
From the leading order equations
[the $(t,t)$ and $(r,r)$ components give the
same equation], we have
\begin{eqnarray}
\phi_1&=&\frac{-r_h\pm\sqrt{r_h^2-8\alpha}}{r_h^2+4\alpha},
\label{phi_1}
\\
b_1&=&\frac{2}{r_h\pm\sqrt{r_h^2-8\alpha}},
\label{b_1}
\end{eqnarray}
and
\begin{eqnarray}
3\frac{a_2}{a_1}+\frac{b_2}{b_1}=-\frac{8r_h}{r_h^2+4\alpha}.
\label{38}
\end{eqnarray}
Only in the upper ``$+$'' sign case,
$b_1$ is well behaved in the $\alpha \to 0$ limit.
Black string solutions are found only in this branch,
and one can easily confirm that
the perturbative solution indeed chooses this branch.
Since $\phi_1$ and $b_1$ must be real,
for given $\alpha$ there is a minimum horizon size:
\begin{eqnarray}
r_h \geq\sqrt{8\alpha}.
\end{eqnarray}
From the next-to-leading order equations and Eq.~(\ref{38})
we can determine the second derivatives of the metric functions at the horizon.
For the $+$ sign case they are given by
\begin{eqnarray}
\frac{a_2}{a_1}=
-\frac{3r_h}{r_h^2+4\alpha}
-\frac{1}{2}\left[
\frac{1}{r_h}-\frac{3r_h^2+4\alpha}{(r_h^2+4\alpha)\sqrt{r_h^2-8\alpha}}
\right],
\\
\frac{b_2}{b_1}=
\frac{r_h}{r_h^2+4\alpha}
+\frac{3}{2}\left[
\frac{1}{r_h}-\frac{3r_h^2+4\alpha}{(r_h^2+4\alpha)\sqrt{r_h^2-8\alpha}}
\right],
\quad
\\
\phi_2=\frac{4\alpha}{r_h(r_h^2+4\alpha)^2}
\hspace{45mm}\\
\times
\frac{r_h(3r_h^2-4\alpha)\sqrt{r_h^2-8\alpha} +(3r_h^2+4\alpha)(r_h^2-4\alpha)}
{r_h(r_h^2-8\alpha)+(r_h^2-4\alpha)\sqrt{r_h^2-8\alpha}}.\nonumber
\end{eqnarray}
It can be seen from these equations that
taking $r_h\to\sqrt{8\alpha}$,
the coefficients $a_2,~b_2$, and $\phi_2$ diverge,
which leads to a singular solution
in the minimum horizon size limit.

Repeating the same procedure,
we obtain all the coefficients $b_i$ and $\phi_i$ ($i=1,2,...$)
and the ratio $a_j/a_1$ ($j=2,3,...$),
but $a_1$ itself and $\Phi_h$ are left undetermined.
The values of these two coefficients
are shifted by rescaling $t$ and $y$ coordinates as
$t\to \lambda_t t$ and $y\to \lambda_y y$ with constant $\lambda_t$ and $\lambda_y$
and are to be fixed so that
$e^{2A(\infty)}=e^{2\Phi(\infty)}=1$.
Thus, the only parameter we can choose freely at the regular horizon
is the value of the horizon radius $r_h$,
and this degree of freedom corresponds
to one freely chosen charge at infinity.

Now we mention
the implication of the minimum mass.
Let us consider a length scale $l$ where
the Einstein-Hilbert term and the Gauss-Bonnet term become comparable.
Since $\car \sim M/r^3$ and ${\cal L}_{GB}\sim \car^2$,
$l$ is roughly given by $\sim (\alpha M)^{1/3}$.
For large mass we have $l \ll r_h\sim M$,
so that $\car \gg \alpha {\cal L}_{GB}$ outside the horizon.
Even for the minimum mass, we have $M_0\sim \alpha^{1/2}$
and thus the scale $l$ is of order the horizon size:
$l\sim \alpha^{1/2}\sim\sqrt{8\alpha}$.
In other words, \textit{the Gauss-Bonnet term
can never dominate the Einstein-Hilbert term outside the horizon.}

\begin{figure}[t]
  \begin{center}
    \includegraphics[keepaspectratio=true,height=50mm]{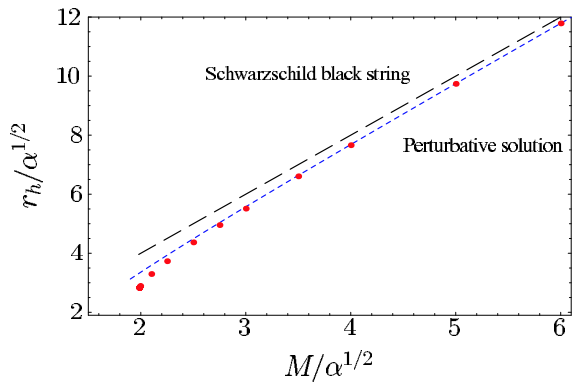}
  \end{center}
  \caption{(color online) Horizon radius as a function of $M$ (red circles).
  Blue dashed line indicates the horizon radius of the perturbative solution,
  $r_h=2M-23\alpha/(18M)$, which approximates the regular black string quite well
  for $M^2\gg \alpha$.}
  \label{fig:mass_vs_horizon.eps}
\end{figure}

\begin{figure}[tb]
  \begin{center}
    \includegraphics[keepaspectratio=true,height=50mm]{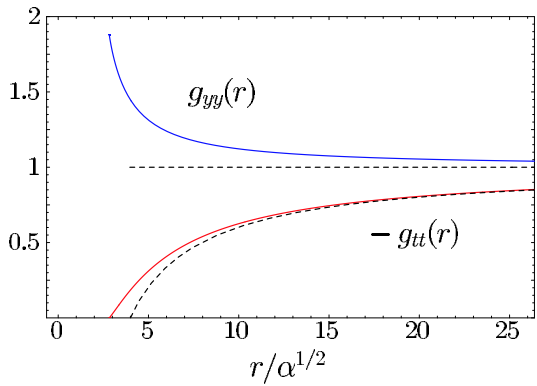}
  \end{center}
  \caption{(color online) Quantitative difference between the minimum mass
  regular black string in Einstein-Gauss-Bonnet gravity
  (red and blue solid lines) and
  the Schwarzschild black string in Einstein gravity with the same mass
  (dashed lines).}
  \label{fig:bs.eps}
\end{figure}

The horizon radius is turned out to be monotonically decreasing with decreasing mass,
and reaches the minimum at $M=M_0$ (Fig.~\ref{fig:mass_vs_horizon.eps}).
We numerically find
$M_0\simeq 1.9843\alpha^{1/2}$ and $Q_c(M_0)\simeq 0.48930\alpha^{1/2}$.
For the minimum mass black string,
the quantitative difference
from the Schwarzschild black string in Einstein gravity with
the same mass stands out well (Fig.~\ref{fig:bs.eps}).

We can relate the expansion coefficients and the asymptotic charges
by using the field equation
${\cal G}_t^{~t}-{\cal G}_y^{~y}=\alpha({\cal H}_t^{~t}-{\cal H}_y^{~y})/2$.
This combination reduces to the total derivative form
\begin{eqnarray*}
&&\frac{d}{dr}\left[
r^2e^{A-B}\left(e^{\Phi}\right)'-r^2e^{\Phi-B}\left(e^{A}\right)'
\right]
\nonumber\\
&&=4\alpha\frac{d}{dr}\left\{
\left(e^{-3B}-e^{-B}\right)\left[
e^A\left(e^{\Phi}\right)'-e^{\Phi}\left(e^{A}\right)'
\right]\right\},
\end{eqnarray*}
and hence
\begin{eqnarray}
&&{\cal F}(r):=r^2e^{\Phi-B}\left(e^{A}\right)'-
r^2e^{A-B}\left(e^{\Phi}\right)'
\nonumber\\
&&\quad+4\alpha\left\{
\left(e^{-3B}-e^{-B}\right)\left[
e^A\left(e^{\Phi}\right)'-e^{\Phi}\left(e^{A}\right)'
\right]\right\},\qquad
\end{eqnarray}
is a constant.
Evaluating this at the horizon and at infinity,
we obtain
\begin{eqnarray}
M+Q_c(M)=\frac{1}{2}(r_h^2+4\alpha) e^{\Phi_h}\sqrt{a_1b_1}.
\end{eqnarray}

\begin{figure}[t]
  \begin{center}
    \includegraphics[keepaspectratio=true,height=160mm]{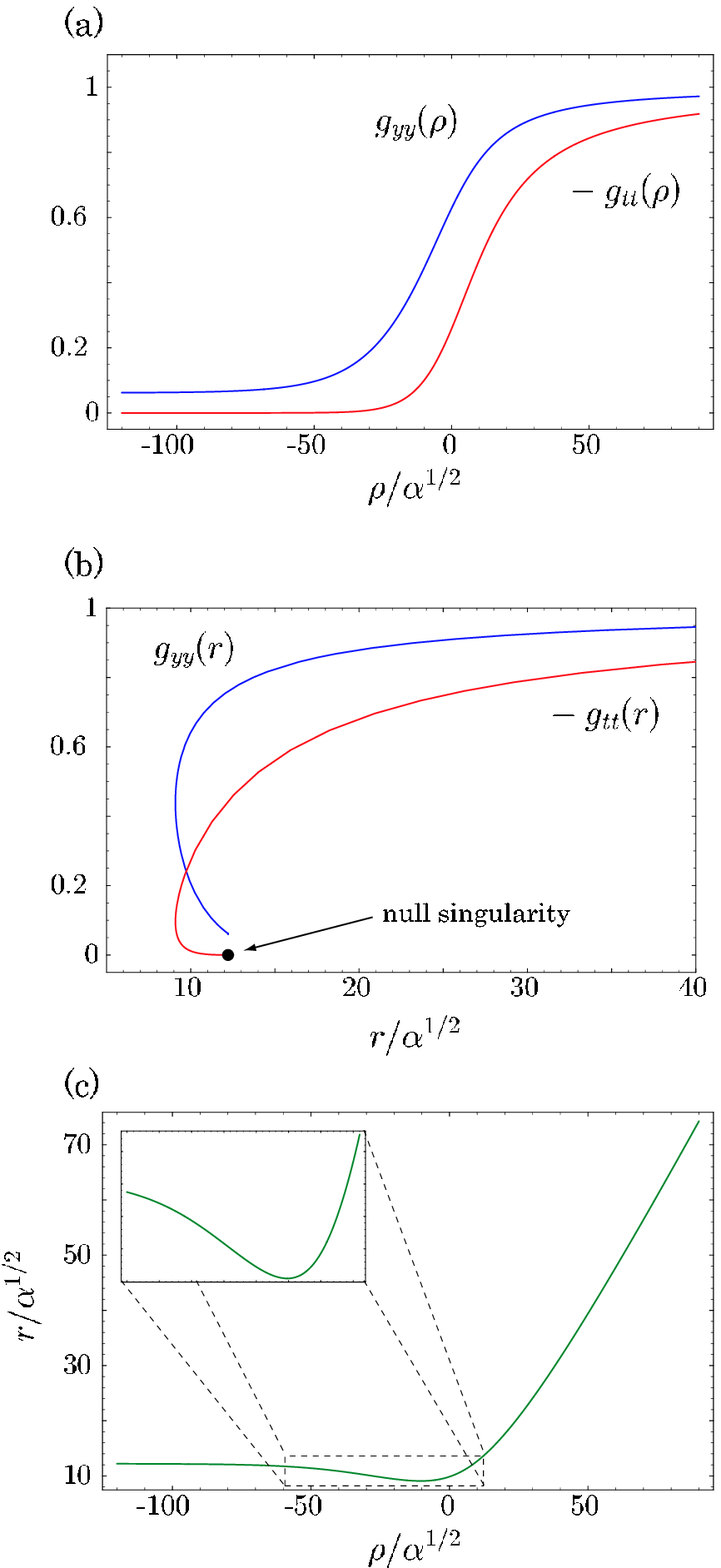}
  \end{center}
  \caption{(color online)
  Metric components $g_{tt}$ (red lines) and $g_{yy}$ (blue lines)
  as functions of the tortoise coordinate $\rho$ (a)
  and the circumferential radius $r$ (b),
  and the relation between $\rho$ and $r$ (c)
  for a black string with
  $M=3\alpha^{1/2}$ and $Q=-\alpha^{1/2}$.
  The two parameters are not fine-tuned.
  The metric components are two-valued functions of $r$, and
  there is a null singularity at the Killing horizon ($g_{tt}=0$).}
  \label{fig:singular.eps}
\end{figure}

\begin{figure}[htbp]
  \begin{center}
    \includegraphics[keepaspectratio=true,height=50mm]{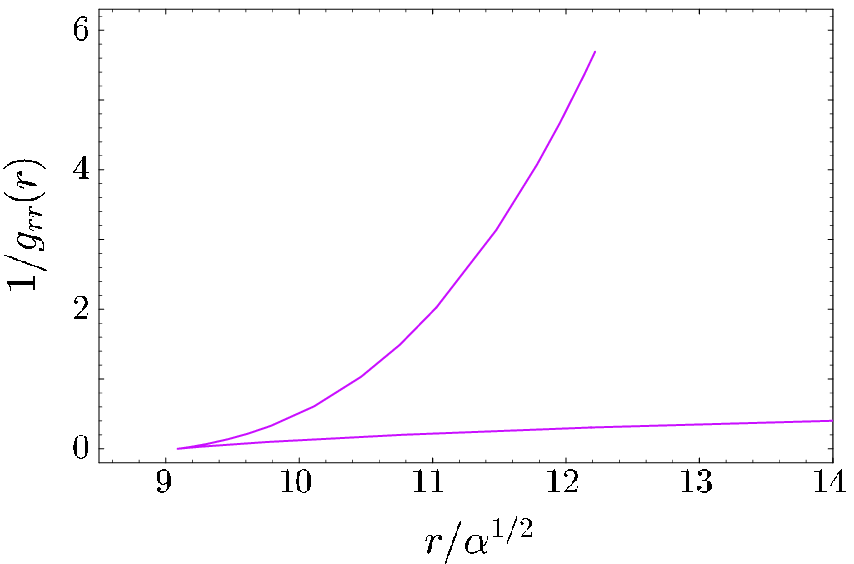}
  \end{center}
  \caption{(color online)
  Behavior of $1/g_{rr}(r)=e^{-2B(r)}$ for $M=3\alpha^{1/2}$
  and $Q=-\alpha^{1/2}$ black string.
  It vanishes at $r=r_m\approx 9.1\alpha^{1/2}$,
  but is finite at $r=r_s\approx 12\alpha^{1/2}$.}
  \label{fig:grr.eps}
\end{figure}

\paragraph{Null singularity}
If one does not impose the fine-tuned relation between
the asymptotic charges, other types of solutions can be found.
For the asymptotic charges satisfying $Q<Q_c(M)$,
the solution shows rather exotic behavior,
as is presented in Figs.~\ref{fig:singular.eps} and~\ref{fig:grr.eps}.
A particular feature is that
the circumferential radius takes minimum at
a finite value of $\rho$ (denoted as $\rho_m$),
where $g_{rr}=(N/\partial_{\rho}r)^2$, $\partial_r g_{tt}$, and $\partial_r g_{yy}$ diverge.

We can find a solution of this class
for $M$ smaller than the minimum mass of the regular black string $M_0$,
which is not contradictory because,
as will be seen shortly, this solution does not have a regular horizon.
The critical value $Q_c(M)$ is not defined for $M<M_0$, but
we can extend the definition of $Q_c(M)$ to the smaller mass region as 
the largest value of $Q$ for which the solution falls in this class. 
Let us denote the extended critical value for $M<M_0$ by $\bar
Q_c(M)$. Namely, the parameter region in which solutions of this class
exist is bounded above by $Q_c$ for $M>M_0$ and by $\bar Q_c$ for $M<M_0$.
For fixed $M$, $\rho_m$ is decreasing with increasing $Q$,
and in the fine-tuned limit we have
\begin{eqnarray}
\lim_{Q\to Q_c}\rho_m\to-\infty.
\end{eqnarray}

Near the minimum circumferential radius, $r=r_m:=r(\rho_m)$,
the metric components behave as
$r-r_m\sim[g_{tt}(r)-g_{tt}(r_m)]^2$, $r-r_m\sim[g_{yy}(r)-g_{yy}(r_m)]^2$,
and $r-r_m\sim (\rho -\rho_m)^2$. 
Noticing these,
we can expand the metric components
as~\cite{Kanti:1995vq, Alexeev:1996vs}
\begin{eqnarray}
&&A(r)=\bar A_m+\bar A_*x^{1/2}+\bar A_1x+\cdots,
\nonumber\\
&&e^{-2B(r)}=\bar b_1x+\cdots,\\
&&\Phi(r)=\bar\Phi_m+\bar \phi_*x^{1/2} +\bar\phi_1x+\cdots,
\nonumber
\end{eqnarray}
where $x=r-r_m$.
The coefficients are constrained by
\begin{eqnarray}
&&\bar b_1=\frac{2}{r_m},
\quad
\bar A_1=-\frac{1}{2}\bar A_{*}^2
-\frac{r_m}{r_m^2+4\alpha},
\nonumber\\
&&\bar \phi_*\bar A_*=\frac{2r_m}{r_m^2+4\alpha},
\quad
\bar\phi_1=-{1\over 2}
\bar \phi_*^2-\frac{r_m}{r_m^2+4\alpha}. 
\label{tenkai-r_m}
\end{eqnarray}
In contrast to the regular horizon case,
there are two free parameters in this expansion.
Namely, given the values of $r_m$ and either $\bar A_*$ or $\bar \phi_*$,
we can determine all the remaining coefficients order by order.
These 2 degrees of freedom correspond to the
two freely chosen asymptotic charges.
Note that $\bar A_m$ and $\bar \Phi_m$,
of which values are shifted by rescaling $t$ and $y$ coordinates,
should be adjusted to satisfy $e^{2A(\infty)}=e^{2\Phi(\infty)}=1$.

Using the expansion~(\ref{tenkai-r_m}), one can easily confirm that
the curvature invariants do not diverge at $r=r_m$,
though, of course, regularity at this point is rather obvious
if one uses the tortoise coordinate.
A similar structure of metric near the minimum circumferential radius
was reported in Ref.~\cite{Kanti:1995vq}
in the framework of dilatonic Gauss-Bonnet gravity~(\ref{dilaton_GB}).
However, since in the analysis of Ref.~\cite{Kanti:1995vq}
they did not use the tortoise coordinate,
and they integrated the field equations from inside toward infinity
using the radial coordinate $r$,
``the structure inside $\rho_m$'' was not recognized.

For $\rho < \rho_m$, the circumferential radius
increases as $\rho$ decreases, until $g_{tt}$ vanishes
at $r(-\infty)=:r_s$.
The metric components
exponentially approach their values at $r=r_s$
as $\rho \to -\infty$:
$g_{tt}\sim e^{c_1\rho}$ and $x:=r_s-r\sim e^{c_2\rho}$
with $c_1,~c_2>0$.
Then, the $(r,r)$ component behaves like
\begin{eqnarray}
e^{-2B}=\left(\frac{\partial_{^\rho}r}{N}\right)^2\sim
e^{(c_1-2c_2)\rho}\sim x^{c_1/c_2-2}.
\end{eqnarray}
We numerically confirmed that
$g_{rr}$ neither vanishes nor diverges at $r=r_s$
(see Fig.~\ref{fig:grr.eps}).
Thus, the only possibility is $c_1/c_2=2$ and
we have $g_{tt}\sim x^2$ at the leading order.
From this argument we obtain
\begin{eqnarray}
&&e^{2A(r)}=\frac{\tilde a_2}{2}x^2-\frac{\tilde a_3}{6}x^3+\cdots,
\nonumber\\
&&e^{-2B(r)}=\tilde b_s-\tilde b_1x+\cdots,\label{expansion_at_r_s}\\
&&\Phi(r)=\tilde\Phi_s-\tilde \phi_1x+\frac{\tilde\phi_2}{2}x^{2} +\cdots.
\nonumber
\end{eqnarray}
The coefficients are constrained by
\begin{eqnarray}
&&\tilde b_1=\frac{r_s}{4\alpha}
\frac{r_s^2+4\alpha+4\alpha\tilde b_s}{r_s^2+4\alpha-4\alpha \tilde b_s},
\label{positivity}
\\
&&\tilde\phi_1=-\frac{2r_s}{r_s^2+4\alpha-12\alpha\tilde b_s},
\\
&&2\frac{\tilde a_3}{\tilde a_2}+3\frac{\tilde b_1}{\tilde b_s}
=-\frac{6r_s(r_s^2+4\alpha-12\alpha\tilde b_s)}
{(r_s^2+4\alpha-4\alpha\tilde b_s)^2},
\\
&&\tilde\phi_2+\tilde \phi_1^2
\nonumber\\&&\quad =
\frac{(1+3\tilde b_s)r_s^2+4\alpha(1-\tilde b_s)(1-3\tilde b_s)}
{\tilde b_s(r_s^2+4\alpha-4\alpha\tilde b_s)(r_s^2+4\alpha-12\alpha\tilde b_s)}.
\end{eqnarray}
Using this expansion,
we can show that
the Kretschmann scalar, $\car_{abcd} \car^{abcd}$, diverges as
\begin{eqnarray}
\car_{abcd} \car^{abcd} \sim {\cal O}\left( \frac{1}{(r_s-r)^2} \right),
\end{eqnarray}
which can be confirmed numerically as well.
Thus, the solution has a null singularity
at the Killing horizon.
Note that here again there are two free parameters, $r_s$ and $\tilde b_s$,
and this is consistent with the analysis near $r=r_m$.

For the same mass black strings, we find that
\begin{eqnarray}
r_h(M, Q_c) < r_s(M, Q),
\end{eqnarray}
namely, the radius of the singular horizon is larger than
that of the event horizon of the regular black string with the same mass.

The constancy of ${\cal F}$ implies that
\begin{eqnarray}
M+Q={\cal F}(r_m)={\cal F}(r_s),
\end{eqnarray}
with
\begin{eqnarray}
{\cal F}(r_m)=\frac{1}{2}(r_m^2+4\alpha) e^{\bar\Phi_m}\sqrt{\bar a_m \bar b_1}
\left( \frac{\bar a_*}{2\bar a_m}-\bar\phi_* \right),
\end{eqnarray}
and
\begin{eqnarray}
{\cal F}(r_s)=(r_s^2+4\alpha-4\alpha \tilde b_s) e^{\tilde\Phi_s}
\sqrt{\frac{\tilde a_2 \tilde b_s}{2}}.
\label{F(r_s)=pos}
\end{eqnarray}
From Eq.~(\ref{positivity})
we see that $r_s^2+4\alpha-4\alpha \tilde b_s>0$
because both $\tilde b_1$ and $\tilde b_s$ are positive,
which yields the positivity of ${\cal F}(r_s)$.
This means that the asymptotic charge $Q$ is restricted in the region
$-M<Q<Q_c(M)$ for this class of solutions.
The limit point $M+Q=0$ can only be achieved
by setting $e^{2\tilde\Phi_s}=g_{yy}(r_s)=0$.
One can easily deduce that
if one chooses the asymptotic charges satisfying $M+Q<0$,
the metric component $g_{yy}$ vanishes at
finite $\rho$, namely,
spacetime has a bubblelike structure.
This is in fact the case, and
such a class of solutions will be discussed in the next subsection.

\paragraph{Naked singularity}
For larger values of the scalar charge, $Q>Q_c(M)$
[or $\bar Q_c(M)$],
we find that $N''(\rho)$, $r''(\rho)$, and $C''(\rho)$
[or $C'(\rho)$] diverge at a finite $\rho$
and the Kretschmann scalar also diverges there;
the solution has a naked singularity.
Thus critical value $Q_c$ and $\bar Q_c$
gives the boundary of
the two classes of solutions,
with a null singularity and with a naked singularity.

\subsection{Bubble dual of black strings}

\begin{figure}[t]
  \begin{center}
    \includegraphics[keepaspectratio=true,height=50mm]{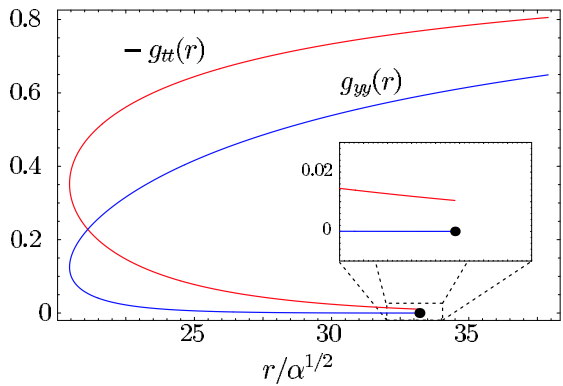}
  \end{center}
  \caption{(color online)
  Behavior of the metric components
  for $M=3\alpha^{1/2}$ and $Q=-6\alpha^{1/2}<-M$.
  The spacetime has a bubblelike structure.
  }
  \label{fig:vanish_gyy.eps}
\end{figure}

As was already mentioned,
we obtain bubble solutions, rather than black strings,
if $M+Q$ is negative (Fig.~\ref{fig:vanish_gyy.eps}).
The behavior of the solutions in the parameter region $M+Q<0$
can be understood
by the following ``dual'' picture~\cite{Sarbach:2004rm}.
Bubble-type spacetime can be constructed from
the black string solutions of the form of~(\ref{r_coordinate})
via a double Wick rotation,
\begin{eqnarray}
t\mapsto iy,\quad y\mapsto it,
\end{eqnarray}
with a simultaneous exchange of the charges,
\begin{eqnarray}
(M,Q) \mapsto (-Q,-M).
\label{exchange}
\end{eqnarray}
The transformation rule~(\ref{exchange})
follows from the asymptotic behavior of $g_{tt}$ and $g_{yy}$
[see Eqs.~(\ref{charge_and_phi}) and~(\ref{asy_M})].
Under this ``duality'' transformation,
any black strings with asymptotic charges $M$ and $Q(>-M)$
are mapped to bubble configurations
with $M'=-Q$ and $Q'=-M$, but now the two charges satisfy
$M'+Q'<0$.
A solution with a null singularity is dual to
a bubblelike solution
that is singular at the point where $g_{yy}$ vanishes,
but this singularity is no longer null.
A naked singularity in the solutions for $Q>Q_c(M)$
remains naked under the duality transformation.
Only for the fine-tuned values of $M$ and $Q$
we obtain a regular bubble solution.

%


\begin{figure}[t]
  \begin{center}
    \includegraphics[keepaspectratio=true,height=50mm]{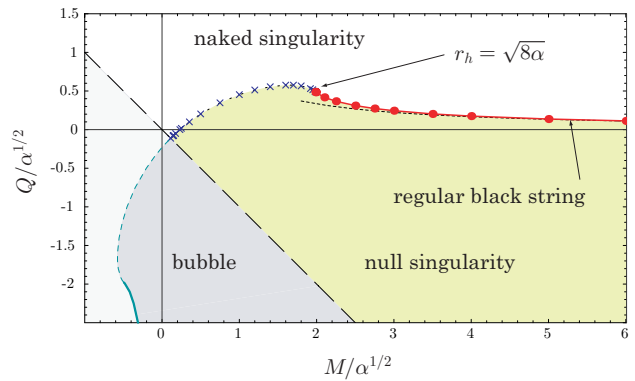}
  \end{center}
  \caption{(color online)
  Classification of solutions with $M$ and $Q$.
  Regular black string solutions can be found only for
  fine-tuned values of $M$ and $Q$ (red solid line),
  and the large one is well approximated by the perturbative solution
  (dashed line just below the red solid one).
  The fine-tuned curve is a boundary of two regions
  in which solutions with a null singularity and a naked singularity,
  respectively, reside.
  This boundary curve can be extended down to
  $M=-Q\approx 0.112\alpha^{1/2}$.
  The nature of the bubble-type solutions
  below the $M+Q=0$ line
  can be understood by the dual picture.
  }
  \label{fig:MvsQ.eps}
\end{figure}


\subsection{Summary}

The results obtained in this section
are summarized in Fig.~\ref{fig:MvsQ.eps}
as a ``phase diagram.''
Although the asymptotic analysis allows
any values of charges $M$ and $Q$,
we have seen that a vast area of the parameter space is covered by
string-type configurations with a singular Killing horizon,
corresponding bubble-type configurations with negative $M+Q$
which is generated by a double Wick rotation,
and solutions with a naked singularity.
We have shown that regular solutions (and their bubble counterparts) form
a one-parameter family whose asymptotic charges are
given by $(M,Q_c(M))$.
We have obtained an approximate solution
that fits quite well the regular black string with $M^2\gg\alpha$
by a perturbative expansion with respect to $\alpha$.
In order for the horizon to be regular,
its radius must be bounded below:~$r_h\geq\sqrt{8\alpha}$,
leading to the existence of the minimum mass black string.

The fine-tuned curve $Q=Q_c(M)$ that generates
the one-parameter family of the regular black strings
can be extended to the small mass region
in the sense that it still represents a boundary of
the classes of the solutions with a null singularity and a naked singularity,
though regular solutions do not exist on the extended curve $Q=\bar Q_c(M)$.
The boundary curve truly terminates at
$(M, Q)\approx(0.112\alpha^{1/2}, -0.012\alpha^{1/2})$,
where it encounters the $M+Q=0$ line.
This line is an invariant set of the transformation $(M, Q)\to(-Q,-M)$,
and solutions reside below this line have a bubblelike structure.

\section{Final remarks}

\begin{figure}[t]
  \begin{center}
    \includegraphics[keepaspectratio=true,height=50mm]{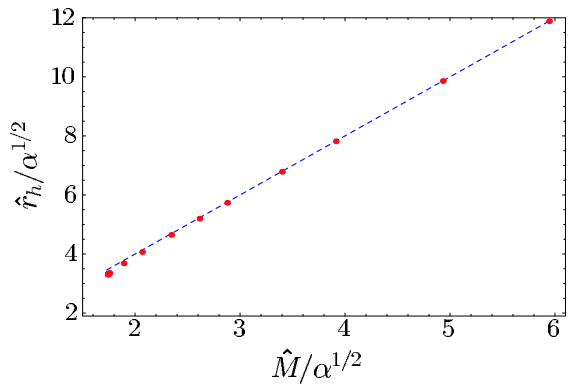}
  \end{center}
  \caption{(color online) Horizon radius in the Einstein frame $\hat r_h$
  as a function of $\hat M$ (red circles).
  Blue dashed line shows the horizon radius of the perturbative solution,
  $\hat r_h =2\hat M$. (It is easy to show that
  $q_{tt}=1-2\hat M/\hat r +{\cal O}(\alpha^2)$ and
  $q_{\hat r\hat r}=(1-2\hat M/\hat r)^{-1} +{\cal O}(\alpha^2)$
  for the perturbative solution~\cite{Mignemi:1992nt}.)}
  \label{fig:M-rh_EF.eps}
\end{figure}

We have seen aspects of black string configurations
in the framework of
five-dimensional Einstein-Gauss-Bonnet gravity.
In the present paper
we focused on static black-string-type metric
respecting spherical symmetry on the transverse hypersurface
and did not consider
(self-gravitating) branes nor a cosmological constant
that warps spacetime.
Consequently our metric ansatz is very simple,
possessing translational invariance in the extra $y$ direction,
but a nontrivial radion part $g_{yy}=e^{2\Phi(r)}$ is assumed,
without which
black-string-type solutions are not allowed
in Einstein-Gauss-Bonnet gravity~\cite{Barcelo:2002wz}.
From a phenomenological point of view,
it would be interesting to generalize
the black string solutions in this paper to afford
brane boundary conditions and a warp factor
due to a bulk cosmological constant.

Stability of solutions is an important issue in black hole physics.
Stability analysis for the four-dimensional black hole solutions
of dilatonic Gauss-Bonnet gravity~(\ref{dilaton_GB})~\cite{Mignemi:1992nt,Kanti:1995vq,Alexeev:1996vs,Torii:1996yi}
was done in Refs.~\cite{Kanti:1997br, Torii:1998gm}. 
Here, following Refs.~\cite{Torii:1998gm, Tamaki:2003ah},
we briefly comment on the stability of
our (regular) black string solution within
translationally invariant and spherically symmetric configurations.
Under the restriction of translationally invariant perturbations
we may work in the reduced action~(\ref{reduced_action}).
In the similar action of dilatonic Gauss-Bonnet gravity~(\ref{dilaton_GB}),
the minimum mass solution does not correspond to 
the end of a sequence of solutions, and
for masses near the
minimum we have two solutions with different horizon radii~\cite{Torii:1998gm}.
With the aid of a catastrophe theory\footnote{It is shown that
stability analysis within {\it linear} perturbations
is consistent with the analysis via a catastrophe theory~\cite{Tamaki:2003ah}.},
one can argue
that an instability mode sets in when the mass 
is reduced to the minimum value.
If we attempt to apply
the same argument to our situation
by seeing whether there exists
such a ``turning point'' in the mass-horizon radius diagram,
we have to use
mass and a horizon radius defined in the Einstein frame~\cite{Tamaki:2003ah},
not $M$ and $r_h$ that we have used throughout this paper.
The Einstein frame metric $q_{\mu\nu}$
is conformally related to the Jordan frame metric $g_{ab}$ via
Eq.~(\ref{einframe}),
\begin{eqnarray}
&&q_{\mu\nu}d\hat x^{\mu} d\hat x^{\nu}
=e^{\Phi}(-e^{2A}dt^2+e^{2B}dr^2+r^2d\Omega^2)
\nonumber\\
&&\quad=-e^{\Phi+2A}dt^2+
\frac{e^{2B}}{\left(1+r\partial_r\Phi/2 \right)^{2}}
d\hat r^2+\hat r^2d\Omega^2,
\quad
\end{eqnarray}
from which mass $\hat M$ and a horizon radius $\hat r_h$
in the Einstein frame can be read off as
\begin{eqnarray}
\hat M =M-\frac{Q}{2},\quad \hat r_h=e^{\Phi(r_h)/2}r_h.
\end{eqnarray}
The relation between $\hat M$ and $\hat r_h$
is plotted in Fig.~\ref{fig:M-rh_EF.eps},
and we see the monotonic behavior, with no turning point in this diagram.
Therefore, it is strongly indicated that
there is no instability under translationally invariant and spherically symmetric,
linear perturbations.
If there arises no 
instability mode down to the minimum mass, the sequence of solutions 
suddenly hits singularity when the mass is reduced, say, by Hawking 
radiation. 
However, another type of instability may arise before reaching 
the minimum mass. 
For example, 
something similar to the Gregory-Laflamme instability of black
strings~\cite{Gregory:1993vy} may happen. 
Since the radius of the extra dimension can be arranged 
to be arbitrary small in the present context,  
the usual criteria for the Gregory-Laflamme instability 
can be avoided easily.
However, the Gauss-Bonnet correction will bring extra terms
to the purterbation equation because of the presence of the
Weyl tensor in the field equations, and such terms
may stabilize or destabilize the black string solutions.
This is an open issue that we hope to return to in a future publication.

\acknowledgments
We thank Makoto Yoshikawa for helpful discussions. 
This work is supported in part 
by Monbukagakusho Grant-in-Aid Nos.
16740141 and 14047212, Inamori Foundation, 
and 21COE program at Kyoto university
``Center for Diversity and Universality in Physics''. 

\newpage


\appendix

\begin{widetext}
\section{Non-zero components of the Einstein and Lanczos tensors}

\begin{itemize}

\item
The Einstein and Lanczos tensors in the coordinate system of~(\ref{r_coordinate}):
\begin{eqnarray}
&&{\cal G}_t^{~t}=\frac{e^{-2B}}{r^2}\left[
r^2\Phi''+2r\Phi'+r^2(\Phi')^2-r^2B'\Phi'-2rB'+1-e^{2B}
\right],
\label{ein_tt}
\\
&&{\cal G}_r^{~r}=\frac{e^{-2B}}{r^2}\left[
r^2A'\Phi'+2r(A'+\Phi')+1-e^{2B}
\right],
\\
&&{\cal G}_y^{~y}=\frac{e^{-2B}}{r^2}\left[
r^2A''+2rA'+r^2(A')^2-r^2A'B'-2rB'+1-e^{2B}
\right],
\\
&&{\cal G}_{\theta}^{~\theta}={\cal G}_{\varphi}^{~\varphi}=
\frac{e^{-2B}}{r^2} [
r^2A''+2rA'+r^2(A')^2-r^2A'B'
\nonumber\\
&&\quad \quad \quad+r^2\Phi''+2r\Phi'+r^2(\Phi')^2-r^2B'\Phi'
-r(A'+\Phi'+B')+r^2A'\Phi'
],
\end{eqnarray}
and
\begin{eqnarray}
&&\frac{1}{8}{\cal H}_t^{~t}=\frac{e^{-2B}}{r^2}
\left\{
(1-3e^{-2B})B'\Phi'-(1-e^{-2B})\left[(\Phi')^2+\Phi''\right]
\right\},
\\
&&\frac{1}{8}{\cal H}_r^{~r}=\frac{e^{-2B}}{r^2}
\left[-(1-3e^{-2B})A'\Phi'\right],
\\
&&\frac{1}{8}{\cal H}_y^{~y}=\frac{e^{-2B}}{r^2}
\left\{
(1-3e^{-2B})A'B'-(1-e^{-2B})\left[(A')^2+A''\right]
\right\},
\\
&&\frac{1}{8}{\cal H}_{\theta}^{~\theta}=\frac{1}{8}{\cal H}_{\varphi}^{~\varphi}=
\frac{e^{-4B}}{r}\left\{
\Phi'\left[A''+(A')^2\right]
+A'\left[\Phi''+(\Phi')^2\right]
-3A'B'\Phi'
\right\},
\label{lanc_thth}
\end{eqnarray}
where ${}':=\partial_r$.

\item
The Einstein and Lanczos tensors in the coordinate system of~(\ref{kame}):
\begin{eqnarray}
&&N^2{\cal G}_t^{~t}=
\frac{C''}{C}-\frac{N'C'}{NC}+2\frac{r''}{r}-2\frac{N'r'}{Nr}
+2\frac{C'r'}{Cr}+\left(\frac{r'}{r}\right)^2-\frac{N^2}{r^2},
\\
&&N^2{\cal G}_{\rho}^{~\rho}=
\frac{N'C'}{NC}+2\frac{N'r'}{Nr}+2\frac{C'r'}{Cr}
+\left(\frac{r'}{r}\right)^2-\frac{N^2}{r^2},
\\
&&N^2{\cal G}_y^{~y}=
\frac{N''}{N}-\left(\frac{N'}{N}\right)^2+2\frac{r''}{r}
+\left(\frac{r'}{r}\right)^2-\frac{N^2}{r^2},
\\
&&N^2{\cal G}_{\theta}^{~\theta}=N^2{\cal G}_{\varphi}^{~\varphi}=
\frac{N''}{N}-\left(\frac{N'}{N}\right)^2+\frac{C''}{C}
+\frac{r''}{r}+\frac{C'r'}{Cr},
\end{eqnarray}
and
\begin{eqnarray}
&&\frac{N^2}{8}{\cal H}_t^{~t}=
-\frac{1}{r^2}\left(\frac{C''}{C}-\frac{N'C'}{NC}\right)
+\frac{1}{N^2}\left[
\frac{C''}{C}\left(\frac{r'}{r}\right)^2+2\frac{C'r'}{Cr}\frac{r''}{r}
-3\frac{N'C'}{NC}\left(\frac{r'}{r}\right)^2
\right],
\\
&&\frac{N^2}{8}{\cal H}_{\rho}^{~\rho}=
-\frac{1}{r^2}\frac{N'C'}{NC}+\frac{3}{N^2}\frac{N'C'}{NC}\left(\frac{r'}{r}\right)^2,
\\
&&\frac{N^2}{8}{\cal H}_y^{~y}=
-\frac{1}{r^2}\left[\frac{N''}{N}-\left(\frac{N'}{N}\right)^2\right]
+\frac{1}{N^2}\left[
\frac{N''}{N}\left(\frac{r'}{r}\right)^2+2\frac{N'r'}{Nr}\frac{r''}{r}
-3\left(\frac{N'}{N}\right)^2\left(\frac{r'}{r}\right)^2
\right],
\\
&&\frac{N^2}{8}{\cal H}_{\theta}^{~\theta}=
\frac{N^2}{8}{\cal H}_{\varphi}^{~\varphi}=\frac{1}{N^2}\left[
\frac{N''C'r'}{NCr}+\frac{N'C''r'}{NCr}+\frac{N'C'r''}{NCr}
-3\frac{C'r'}{Cr}\left(\frac{N'}{N}\right)^2
\right],
\end{eqnarray}
where ${}':=\partial_{\rho}$.
\end{itemize}
\end{widetext}





\begin{thebibliography}{99}




\bibitem{Emparan:2001wn}
R.~Emparan and H.~S.~Reall,
Phys.\ Rev.\ Lett.\  {\bf 88}, 101101 (2002)
[arXiv:hep-th/0110260].

\bibitem{Myers:1986un}
R.~C.~Myers and M.~J.~Perry,
Annals Phys.\  {\bf 172}, 304 (1986).

\bibitem{Wiseman:2002zc}
T.~Wiseman,
Class.\ Quant.\ Grav.\  {\bf 20}, 1137 (2003)
[arXiv:hep-th/0209051].

\bibitem{Kudoh:2004hs}
H.~Kudoh and T.~Wiseman,
``Connecting black holes and black strings,''
arXiv:hep-th/0409111.


\bibitem{Gregory:1993vy}
R.~Gregory and R.~Laflamme,
Phys.\ Rev.\ Lett.\  {\bf 70} (1993) 2837
[arXiv:hep-th/9301052].


\bibitem{Kodama:2003jz}
H.~Kodama and A.~Ishibashi,
Prog.\ Theor.\ Phys.\  {\bf 110}, 701 (2003)
[arXiv:hep-th/0305147].

\bibitem{Ishibashi:2003ap}
A.~Ishibashi and H.~Kodama,
Prog.\ Theor.\ Phys.\  {\bf 110}, 901 (2003)
[arXiv:hep-th/0305185].

\bibitem{Kodama:2003kk}
H.~Kodama and A.~Ishibashi,
Prog.\ Theor.\ Phys.\  {\bf 111}, 29 (2004)
[arXiv:hep-th/0308128].



\bibitem{Emparan:2003sy}
R.~Emparan and R.~C.~Myers,
JHEP {\bf 0309}, 025 (2003)
[arXiv:hep-th/0308056].

\bibitem{Dotti:2004sh}
G.~Dotti and R.~J.~Gleiser,
``Gravitational instability of Einstein-Gauss-Bonnet black holes under tensor
mode perturbations,''
arXiv:gr-qc/0409005.



\bibitem{Chamblin:1999by}
A.~Chamblin, S.~W.~Hawking and H.~S.~Reall,
Phys.\ Rev.\ D {\bf 61}, 065007 (2000)
[arXiv:hep-th/9909205].

\bibitem{Kudoh:2003xz}
H.~Kudoh, T.~Tanaka and T.~Nakamura,
Phys.\ Rev.\ D {\bf 68}, 024035 (2003)
[arXiv:gr-qc/0301089].

\bibitem{Karasik:2003tx}
D.~Karasik, C.~Sahabandu, P.~Suranyi and L.~C.~R.~Wijewardhana,
Phys.\ Rev.\ D {\bf 69}, 064022 (2004)
[arXiv:gr-qc/0309076].


\bibitem{Giddings:2001bu}
S.~B.~Giddings and S.~Thomas,
Phys.\ Rev.\ D {\bf 65}, 056010 (2002)
[arXiv:hep-ph/0106219].

\bibitem{Dimopoulos:2001hw}
S.~Dimopoulos and G.~Landsberg,
Phys.\ Rev.\ Lett.\  {\bf 87}, 161602 (2001)
[arXiv:hep-ph/0106295].




\bibitem{Seahra:2004fg}
S.~S.~Seahra, C.~Clarkson and R.~Maartens,
``Detecting extra dimensions with gravity wave spectroscopy,''
arXiv:gr-qc/0408032.


\bibitem{Mignemi:1992nt}
S.~Mignemi and N.~R.~Stewart,
Phys.\ Rev.\ D {\bf 47}, 5259 (1993)
[arXiv:hep-th/9212146].


\bibitem{Kanti:1995vq}
P.~Kanti, N.~E.~Mavromatos, J.~Rizos, K.~Tamvakis and E.~Winstanley,
Phys.\ Rev.\ D {\bf 54}, 5049 (1996)
[arXiv:hep-th/9511071].

\bibitem{Alexeev:1996vs}
S.~O.~Alexeev and M.~V.~Pomazanov,
Phys.\ Rev.\ D {\bf 55}, 2110 (1997)
[arXiv:hep-th/9605106].


\bibitem{Torii:1996yi}
T.~Torii, H.~Yajima and K.~i.~Maeda,
Phys.\ Rev.\ D {\bf 55}, 739 (1997)
[arXiv:gr-qc/9606034].



\bibitem{gaussbonnet}
  J.~E.~Kim, B.~Kyae and H.~M.~Lee,
  Nucl.\ Phys.\ B {\bf 582}, 296 (2000)
  [Erratum-ibid.\ B {\bf 591}, 587 (2000)]
  [arXiv:hep-th/0004005],
  J.~E.~Kim and H.~M.~Lee,
  Nucl.\ Phys.\ B {\bf 602}, 346 (2001)
  [Erratum-ibid.\ B {\bf 619}, 763 (2001)]
  [arXiv:hep-th/0010093],
  K.~A.~Meissner and M.~Olechowski,
  Phys.\ Rev.\ D {\bf 65}, 064017 (2002)
  [arXiv:hep-th/0106203],
  I.~P.~Neupane,
  Phys.\ Lett.\ B {\bf 512}, 137 (2001)
  [arXiv:hep-th/0104226],
  Y.~M.~Cho, I.~P.~Neupane and P.~S.~Wesson,
  Nucl.\ Phys.\ B {\bf 621}, 388 (2002)
  [arXiv:hep-th/0104227],
  I.~P.~Neupane,
  Class.\ Quant.\ Grav.\  {\bf 19}, 5507 (2002)
  [arXiv:hep-th/0106100],
  Y.~M.~Cho and I.~P.~Neupane,
  Int.\ J.\ Mod.\ Phys.\ A {\bf 18}, 2703 (2003)
  [arXiv:hep-th/0112227],
  J.~P.~Gregory and A.~Padilla,
  Class.\ Quant.\ Grav.\  {\bf 20}, 4221 (2003)
  [arXiv:hep-th/0304250].
  
\bibitem{Charmousis:2003ke}
  C.~Charmousis, S.~C.~Davis and J.~F.~Dufaux,
  JHEP {\bf 0312}, 029 (2003)
  [arXiv:hep-th/0309083].

\bibitem{gbbh}
  D.~G.~Boulware and S.~Deser,
  Phys.\ Rev.\ Lett.\  {\bf 55}, 2656 (1985),
  R.~G.~Cai,
  Phys.\ Rev.\ D {\bf 65}, 084014 (2002)
  [arXiv:hep-th/0109133].

\bibitem{Konoplya:2004xx}
  R.~Konoplya,
  Phys.\ Rev.\ D {\bf 71}, 024038 (2005)
  [arXiv:hep-th/0410057].


\bibitem{Cvetic:2001bk}
  M.~Cvetic, S.~Nojiri and S.~D.~Odintsov,
  Nucl.\ Phys.\ B {\bf 628}, 295 (2002)
  [arXiv:hep-th/0112045].

\bibitem{Charmousis:2002rc}
C.~Charmousis and J.~F.~Dufaux,
Class.\ Quant.\ Grav.\  {\bf 19}, 4671 (2002)
[arXiv:hep-th/0202107].

\bibitem{Nojiri:2000gv}
  S.~Nojiri and S.~D.~Odintsov,
  JHEP {\bf 0007}, 049 (2000)
  [arXiv:hep-th/0006232],
  S.~Nojiri, S.~D.~Odintsov and S.~Ogushi,
  Phys.\ Rev.\ D {\bf 65}, 023521 (2002)
  [arXiv:hep-th/0108172].


\bibitem{gbcosmology}
  J.~E.~Lidsey and N.~J.~Nunes,
  Phys.\ Rev.\ D {\bf 67}, 103510 (2003)
  [arXiv:astro-ph/0303168],
  G.~Kofinas, R.~Maartens and E.~Papantonopoulos,
  JHEP {\bf 0310}, 066 (2003)
  [arXiv:hep-th/0307138].



\bibitem{Deruelle:2003ck}
N.~Deruelle and J.~Madore,
``On the quasi-linearity of the Einstein- 'Gauss-Bonnet' gravity field
equations,''
arXiv:gr-qc/0305004.



\bibitem{Barcelo:2002wz}
C.~Barcelo, R.~Maartens, C.~F.~Sopuerta and F.~Viniegra,
Phys.\ Rev.\ D {\bf 67}, 064023 (2003)
[arXiv:hep-th/0211013].

\bibitem{Sarbach:2004rm}
O.~Sarbach and L.~Lehner,
``Critical bubbles and implications for critical black strings,''
arXiv:hep-th/0407265.


\bibitem{Kanti:1997br}
P.~Kanti, N.~E.~Mavromatos, J.~Rizos, K.~Tamvakis and E.~Winstanley,
Phys.\ Rev.\ D {\bf 57}, 6255 (1998)
[arXiv:hep-th/9703192].


\bibitem{Torii:1998gm}
T.~Torii and K.~i.~Maeda,
Phys.\ Rev.\ D {\bf 58}, 084004 (1998).

\bibitem{Tamaki:2003ah}
T.~Tamaki, T.~Torii and K.~i.~Maeda,
Phys.\ Rev.\ D {\bf 68}, 024028 (2003).


\end{thebibliography}
\end{document}